\documentclass[a4paper,aps,superscriptaddress,floatfix,nofootinbib,twocolumn,showkeys]{revtex4-2}

\usepackage[svgnames,usenames,dvipsnames]{xcolor}
\usepackage{lipsum}
\usepackage{colortbl}
 \usepackage{booktabs}
 \usepackage{multirow}
 \usepackage{graphicx,bigdelim}

\usepackage{comment}
\usepackage{pgfplots}
\usepackage{pgfplotstable}
\usetikzlibrary{pgfplots.groupplots,arrows,automata,positioning}
\usepgfplotslibrary{dateplot,fillbetween}
\usepgfplotslibrary{dateplot}
\pgfplotsset{compat=newest}
\usetikzlibrary{spy}

\pgfplotsset{every axis/.append style={
                    label style={font=\sffamily\large},
                    tick label style={font=\sffamily\scriptsize},
                    title style = {font=\sffamily\large},
                    ylabel near ticks,
                    y label style={font=\sffamily\large},
                    xlabel near ticks,
                    x label style={font=\sffamily\large},
                    legend cell align={left},
                    legend style={draw=none, font=\sffamily\normalsize},
                    },
                    legend image code/.code={
                    \draw[mark repeat=2,mark phase=2]
                        plot coordinates {
                        (0cm,0cm)
                        (0.15cm,0cm)        
                        (0.3cm,0cm)         
                        };%
                    },
                    boxplot/hide outliers/.code={
                        \def\pgfplotsplothandlerboxplot@outlier{}%
                    }
                    }
\pgfplotsset{compat=newest}  
\pgfplotsset{
    jitter/.style={
        x filter/.code={\pgfmathparse{\pgfmathresult+(rnd-.5)*#1}}
    },
    jitter/.default=0.1
}

\def\addlegendimage{\csname pgfplots@addlegendimage\endcsname}

\newcommand{\ie}{\textit{i.e.}}
\newcommand{\etal}{\textit{et al.}}
\newcommand{\para}[1]{\textbf{#1}.\quad}
\usepackage{hyperref}
\usepackage{amsmath}
\usepackage{nicematrix}

\hypersetup{
    colorlinks,
    linkcolor={red!50!black},
    citecolor={blue!50!black},
    urlcolor={blue!80!black}
}

\begin{document}

\title{Consistency pays off in science}

\author{\c{S}irag Erkol}
\affiliation{Center for Complex Networks and Systems Research, Luddy School of Informatics, Computing, and Engineering, Indiana University Bloomington, Indiana 47408, USA}

\author{Satyaki Sikdar}
\affiliation{Center for Complex Networks and Systems Research, Luddy School of Informatics, Computing, and Engineering, Indiana University Bloomington, Indiana 47408, USA}

\author{Filippo Radicchi}
\affiliation{Center for Complex Networks and Systems Research, Luddy School of Informatics, Computing, and Engineering, Indiana University Bloomington, Indiana 47408, USA}

\author{Santo Fortunato}
\affiliation{Center for Complex Networks and Systems Research, Luddy School of Informatics, Computing, and Engineering, Indiana University Bloomington, Indiana 47408, USA}
\affiliation{Indiana University Network Science Institute (IUNI), Bloomington, Indiana 47408, USA}

\begin{abstract}

The exponentially growing number of scientific papers stimulates a discussion on the interplay between quantity and quality in science. In particular, one may wonder which publication strategy may offer more chances of success: publishing lots of papers, producing a few hit papers, or something in between. Here we tackle this question by studying the scientific portfolios of Nobel Prize laureates. A comparative analysis of different citation-based indicators of individual impact suggests that the best path to success may rely on consistently producing high-quality work. Such a pattern is especially rewarded by a new metric, the $E$-index, which identifies excellence better than state-of-the-art measures.

\end{abstract}

\maketitle

\section{Introduction}

The number of scientific papers has been growing exponentially for over a century~\cite{dong17,fortunato18}. The number of papers per author has been relatively stable for a long time, but it has been increasing over the past decades~\cite{dong17}, favored by the growing tendency of scientists to work in teams~\cite{wuchty07}. 

Such increased productivity is incentivized by career evaluation criteria that typically reward large outputs, making scientists less risk averse when choosing research directions~\cite{franzoni2017academic}. This, however,
may come at the expense of the quality of research outcomes \cite{bornmann2019productivity, sunahara2021association}. Indeed, it has been shown that the exponential growth of the number of publications corresponds to a much slower increase in the number of new or disruptive ideas~\cite{milojevic15, chu2021slowed}.

However, while scholars should focus on quality, it is unclear whether it is more rewarding to pursue rare hit papers, have a consistent track record of valuable outputs, or be in between these scenarios.
Analyzing the careers of arguably the most successful class of scientists, Nobel Prize laureates, may help address this issue. In particular, we would like to check if there is a dominant path to success in the careers of such illustrious scholars. 

To that effect, we consider a broad range of evaluation metrics that reward one-hit wonders alongside those that favor a consistent production of high-quality research and investigate their effectiveness in identifying Nobelists from within a more extensive set of similarly productive scientists.
We find that the best-performing metrics are indeed the ones that prioritize a consistent stream of high-quality research.

The rest of this article is organized as follows. We first describe the data collection and curation in Sec.~\ref{sec:data}. Then, we briefly review some popularly adopted impact metrics and introduce two new ones. In Sec.~\ref{sec:experiments}, we describe and discuss the two sets of experiments we used to check which of the two competing scenarios is more common. Finally, we give our conclusions in Sec.~\ref{sec:conclusion}.

\section{Methods} \label{sec:data}

\subsection{Data}

We consider three fields in which the Nobel Prize is awarded: Physics, Chemistry, and Physiology or Medicine (abbreviated henceforth as Medicine). 

The publication records of scientists are obtained from two sources. 
For Nobelists, we use the hand-curated dataset with explicit annotations for prize-winning papers~\cite{li2019dataset}. 
As a baseline, we consider scientists with verified Google Scholar (GS) profiles tagged with either Physics, Chemistry, Physiology, or Medicine as of May 2021. 

We use the 2017 version of the Web of Science (WoS) database to compile the citation statistics of the articles. We rely on gathering data from different sources on purpose, as WoS and GS complement each other well. GS offers the possibility of obtaining accurate publication records of individual scientists without the need to perform name disambiguation~\cite{radicchi2013analysis}. WoS lets us reconstruct the citation history of individual papers. Both ingredients are necessary for the type of analysis that we perform in this paper. 
 
We adopt a similar methodology as Sinatra \etal~\cite{sinatra2016quantifying} to match papers across databases.  
Given a paper $\hat p$ written by author $a$ in GS, we list the papers $P_a$ in WoS authored by people with the same last name as $a$.
From $P_a$, we select the paper $p$ with the highest normalized Levenshtein similarity between the corresponding paper titles~\cite{levenshtein}. 
We consider it a successful match only if the similarity exceeds 90\%. Otherwise, we discard $\hat{p}$ from further analysis. 
Following this procedure, we could match 78.1\% of papers by Nobelists and 49.6\% of papers by baseline scientists, respectively. For our analysis, we only consider scientists who published their first paper after 1960 and have a portfolio with at least ten papers. Detailed statistics are provided in Table~\ref{table:dataset}.

\begin{table}[ht]
    \caption{Number of scientists in each category and field.} 
    \begin{tabular}{@{} lrrr @{}}
        \toprule
        \textbf{Category} & \textbf{Physics} & \textbf{Chemistry} & \textbf{Medicine} \\
        \midrule
        Nobelists & 55 & 51 & 56 \\
        Baseline scientists & 4,081 & 3,330 & 2,715 \\
        \bottomrule
    \end{tabular}
    \label{table:dataset}
\end{table}

\subsection{Metrics}  \label{sec:metrics}

Let us consider a portfolio $\mathcal{P} = \{c_1, \ldots, c_N\}$ of $N = |\mathcal{P}|$ papers that collectively receive $C_\text{tot}$ citations, \ie, $C_\text{tot} = \sum_{i}^{N} c_i$. We consider the following metrics:

\begin{description}
    \item[$N$] total number of papers.

    \item[$C_\textmd{tot}$] total number of citations.

    \item[$C_\textmd{avg}$]  average number of citations, \ie, $C_\text{avg}(\mathcal{P}) = \tfrac{C_\text{tot}}{N}$.

    \item[$C_\textmd{max}$] citations received by the most cited paper, \ie, $C_\text{max}(\mathcal{P}) = {\max} \{c_1, \cdots, c_N\}$.

    \item[$H$] H-index, \ie, the largest number $H$ of the top-cited papers with at least $H$  citations~\cite{hirsch2005index}.

    \item[$G$] G-index, \ie, the largest number $G$ of the top-cited papers with at least $G^2$ combined citations~\cite{egghe2006theory}.

    \item[$Q$] Q-index, proposed by Sinatra \etal, 
    $Q(\mathcal{P}) = \exp \left( \frac{1}{\sum_{i=1}^N \, \Theta(c_{10,i})} \,  \sum_{i=1}^N \, \Theta(c_{10,i}) \, \log c_{10,i}  \right)$, up to a constant factor, where $\Theta$ is the Heaviside function, \ie, $\Theta(x) = 1$ if $x > 0$ and $0$ otherwise, and $c_{10,i}$ is the citations gained by paper $i$ within 10 years of publication. We normalize $c_{10,i}$ by dividing it with the average $c_{10}$ of all papers published in the same discipline and year as paper $i$~\cite{sinatra2016quantifying}.

    \item[$\tilde Q$] a variant of the unnormalized $Q$-index, where we use the total number of citations $c_i$ instead of $c_{10,i}$. 
\end{description}

We observe that these measures have their unique preferences for ranking portfolios. Some, like $C_\text{max}$, appear to reward one-hit wonders, while others, like $H$, reward consistency. 
One of the goals of this work is to identify and differentiate Nobelists from baseline scientists. 
Therefore, we argue that we need a new, simple, yet interpretable metric covering the whole portfolio spectrum.

\subsection{Citation Moment and $E$-index}
Given a publication portfolio $\mathcal{P}$, one may consider the following extreme scenarios: 

\begin{itemize}
    \item Citations are equally distributed among the papers, with each paper having $C_\text{tot}/N$ citations.
    
    \item A single paper accounts for all citations.
\end{itemize}

In the first case, there is a sustained production of work of similar quality, while the second represents a one-hit-wonder situation.

\noindent\para{Citation Moment} We propose the \textit{citation moment} $M_{\alpha}$, a new parametric measure that can reward both scenarios, as well as the ones in between, depending on the value of the parameter $\alpha$. It is defined as 

\begin{equation}
    M_\alpha (\mathcal{P})
    = \frac{1}{N} \; \sum_{i=1}^N c_{i}^\alpha \; ,
    \label{eq:alpha_score}
\end{equation}

where $\alpha$ is a real positive number.
We remark that $M_{\alpha}$ is essentially an average of the citation scores of the papers, where the weight of each score is modulated by the exponent $\alpha$. 
We can make the following observations of the behavior of our metric for different values of $\alpha$.

\begin{figure*}
    \centering
    \includegraphics[width=0.97\textwidth]{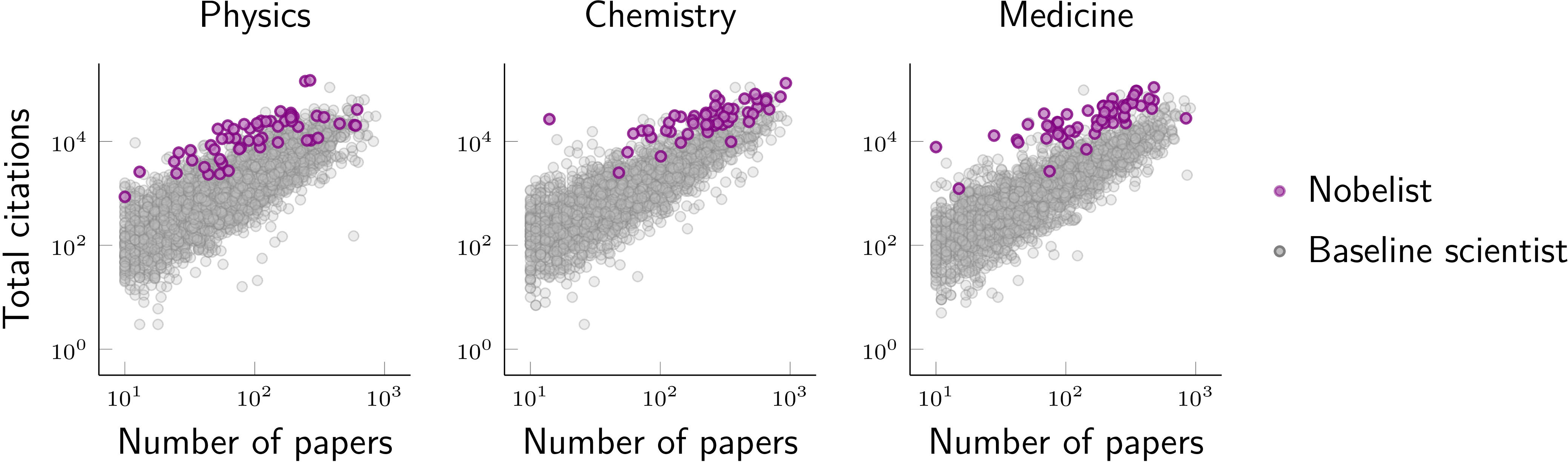}
    \caption{Total number of citations {\it vs.} total number of papers
    for Nobelists (purple dots) and baseline scientists (grey dots).}
    \label{fig:NC}
\end{figure*}

\begin{description}
    \item[$\alpha \rightarrow 0$] $M_{\alpha}$ behaves like $\tilde Q$ as $c^\alpha \approx \log c$, but unlike $\tilde Q$, it accounts for uncited papers.

    \item[$0 < \alpha < 1$]
    $M_{\alpha}$ is higher for balanced portfolios, \ie, ones with a more uniform distribution of citations.

    \item[$\alpha=1$] $M_{\alpha}$ becomes identical to $C_\text{avg}$.

    \item[$\alpha > 1$] $M_{\alpha}$ is higher for unbalanced portfolios.

    \item[$\alpha \rightarrow \infty$] $M_{\alpha}$ closely imitates  $C_\text{max}$.

\end{description}

\noindent\para{$E$-index} We also propose an additional parameter-free measure that, like $M_{\alpha}$, is sensitive to the distribution of citations. 
We call this metric $E$-index, defined as
\begin{equation}
    E(\mathcal{P})
    = - \frac{1}{N}\sum_{i=1}^N c_{i} \log \frac{c_{i}}{C_\text{tot}} \; ,
    \label{eq:entropy}
\end{equation}
which reaches its maximum $C_\text{avg}  \log N$ when citations are distributed equally among papers, favoring authors with large average numbers of citations. In fact, $E(\mathcal{P})$
is just the product of the average number of citations $C_\text{avg}$ and of the Shannon entropy of the citation distribution.

\subsection{Behavior of metrics on stylized portfolios}
To better understand the behavior of the different metrics in our analysis, we consider a portfolio with $n$ cited papers with $C_\text{tot}/n$ citations each and $N-n$ uncited papers. In Table~\ref{table:limiting}, we show the values several key metrics take in this case. 

We see that the citation moment $M_{\alpha}$ (for $\alpha \neq 0,1$), $E$-index, and $G$-index depend on $n$, $N$, and $C_\text{tot}$. The $H$-index and the $\tilde Q$ depend only on the cited papers. So, for example, two portfolios with identical values of $C_\text{tot}$ and $n$ would have the same $H$-index, regardless of the number of uncited papers. Furthermore, even though $G$-index depends on all three parameters, it depends on them in a somehow undesirable way. For example, a portfolio with more uncited papers may have a $G$-index value greater than or equal to the $G$-index of another portfolio with identical $C_\text{tot}$ and $n$ values. Instead, ranking the portfolio with fewer uncited works higher (lower $N-n$), as $M_{\alpha}$ and $E$ would, seems more intuitive. 

\begin{table}[htb]
    \caption{Values of metrics for portfolios with $N$ papers with $C_\text{tot}$ citations, of which $n$ are equally cited and $N - n$ are uncited.}
    \label{table:limiting}
    \centering
    \begin{tabular}{@{}>{\columncolor{white}[0pt][\tabcolsep]}  l c >{\columncolor{white}[\tabcolsep][0pt]}c @{}}
        \toprule
        \textbf{Metric} & \phantom{aaaaaaaaaaaaaaa} & \textbf{Value} \\
        \midrule
        $H$ && $\min \{ \lfloor C_\text{tot}/n \rfloor , n \}$ \\[7pt]
        $G$ && $\min \{ \lfloor \sqrt{C_\text{tot}} \rfloor, \lfloor C_\text{tot}/n \rfloor, N \}$ \\[7pt]
        $\tilde Q$ && $C_\text{tot}/n$ \\[7pt]
        $M_{\alpha}$ && $\frac{C_\text{tot}^\alpha}{N \, n^{\alpha-1}}$ \\ [7pt]
        $E$ && $\frac{C_\text{tot}}{N} \log {n}$ \\
        \bottomrule
    \end{tabular}
\end{table}

\section{Results} \label{sec:experiments}
\begin{figure}[htb!]
    \centering
    \includegraphics[width=0.45\textwidth]{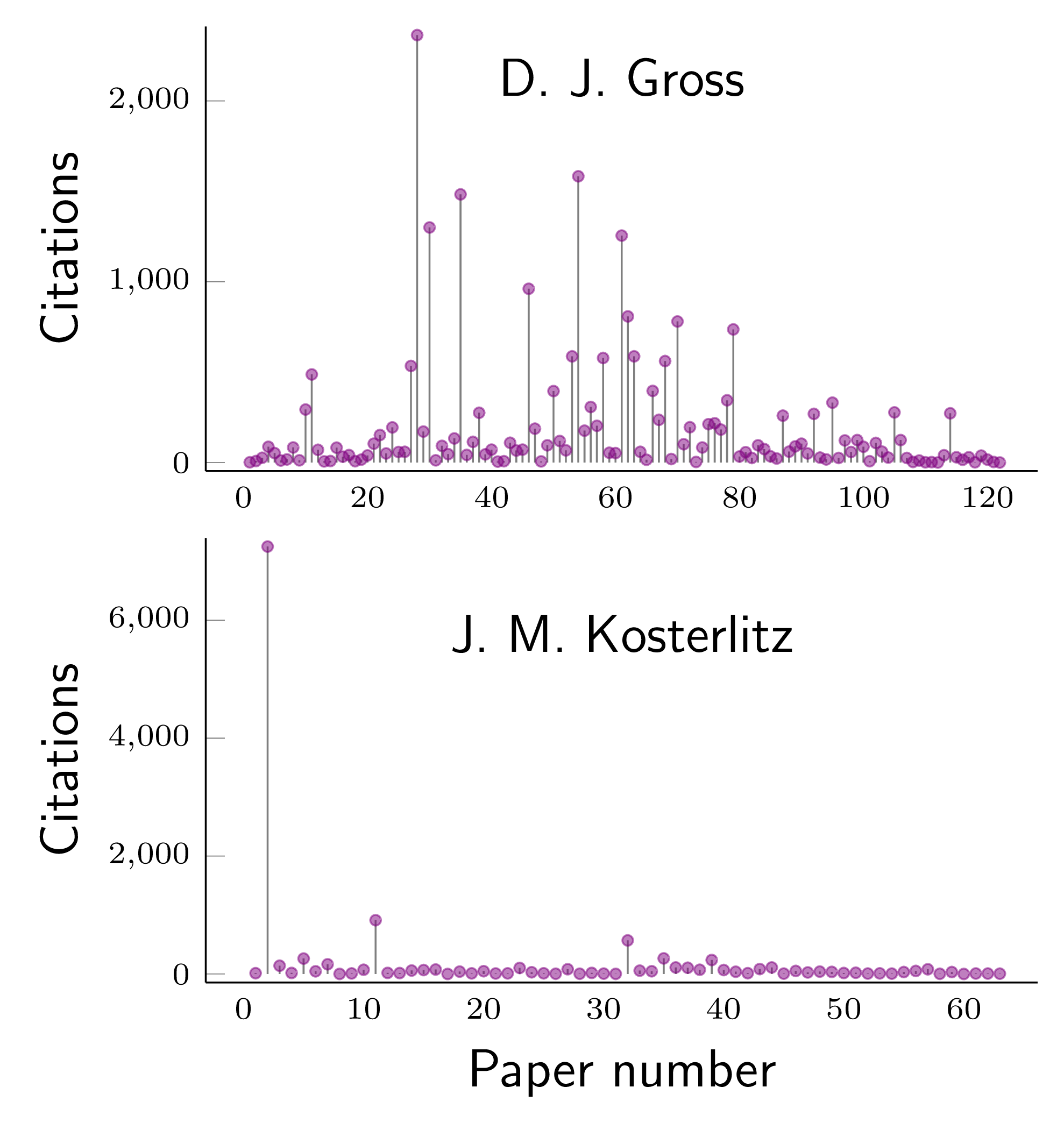}
    \caption{Consistency versus single-hit scenario. On the x-axis, we indicate the temporal sequence of papers, and on the y-axis, the citations accrued by each paper. The two panels show the profiles of D. J. Gross (top) and J. M. Kosterlitz (bottom). The former has a portfolio with multiple highly cited papers, and the latter has one highly cited paper. D. J. Gross: $N=122$, $C_\text{tot}=24,144$, $C_\text{avg}=197.9$, $E=768.6$. J. M. Kosterlitz: $N=63$, $C_\text{tot}=11,688$, $C_\text{avg}=185.5$, $E=348.8$.}
    \label{fig:twonobels}
\end{figure}

In Fig.~\ref{fig:NC}, we plot Nobelists and baseline scientists according to their number of papers and the total number of citations. 
As expected, most Nobelists lie in the top right region, indicating high levels of both productivity and impact. 
However, there appear to be a few Nobelists in the top left, indicating that they only produced a handful of high-impact papers.
To further illustrate this difference, we consider two Nobelists in Physics, David J. Gross (2004) and John M. Kosterlitz (2016), and plot their publication timelines in Fig.~\ref{fig:twonobels}. 
Gross has a consistent production of high-impact
works, while Kosterlitz stands out for having a single big paper. 

We now focus on two tasks: portfolio classification and future Nobelist identification.

\subsection{Portfolio classification}

We test the performance of the metrics in distinguishing the portfolios of Nobelists from those of the baseline scientists. We consider two subtasks which we describe below. We use the area under the precision-recall curve (AUC-PR) in each task as the performance metric. This curve shows the trade-off between precision and recall at different thresholds. Bounded between 0 and 1, higher AUC-PR values indicate better classification performance. For random predictions, AUC-PR is the fraction of positive samples. AUC-PR is better suited for imbalanced datasets than the area under the receiver operating characteristic curve (ROC-AUC)~\cite{saito2015precision}. Results for the ROC-AUC are reported in Appendix \ref{sec:appB} and are consistent with the analysis done using AUC-PR.

\noindent\para{Full} We use the entire portfolio of the scientists described in Sec.~2(a).

\noindent\para{Pre-award} 
We construct the pre-award portfolio of Nobelists, \ie, the set of papers published until the year of the prize-winning paper,  discarding those with fewer than ten papers. We find that 15 (27\%), 28 (55\%), and 22 (39\%) of Nobelists in Physics, Chemistry, and Medicine, respectively, satisfy the above criteria.

Specifically, for a Nobelist who published their first paper in year $y_0$ and wrote their prize-winning article in year $y_p$, we consider the papers published and citations accrued between years $y_0$ and $y_p-1$.
We then pair the Nobelist with 20 baseline scientists who published their first papers around the year $y_0$ and wrote at least ten papers in their careers' first $y_p-y_0$ years.

\begin{figure}[!b]
    \centering
    \includegraphics[width=0.45\textwidth]{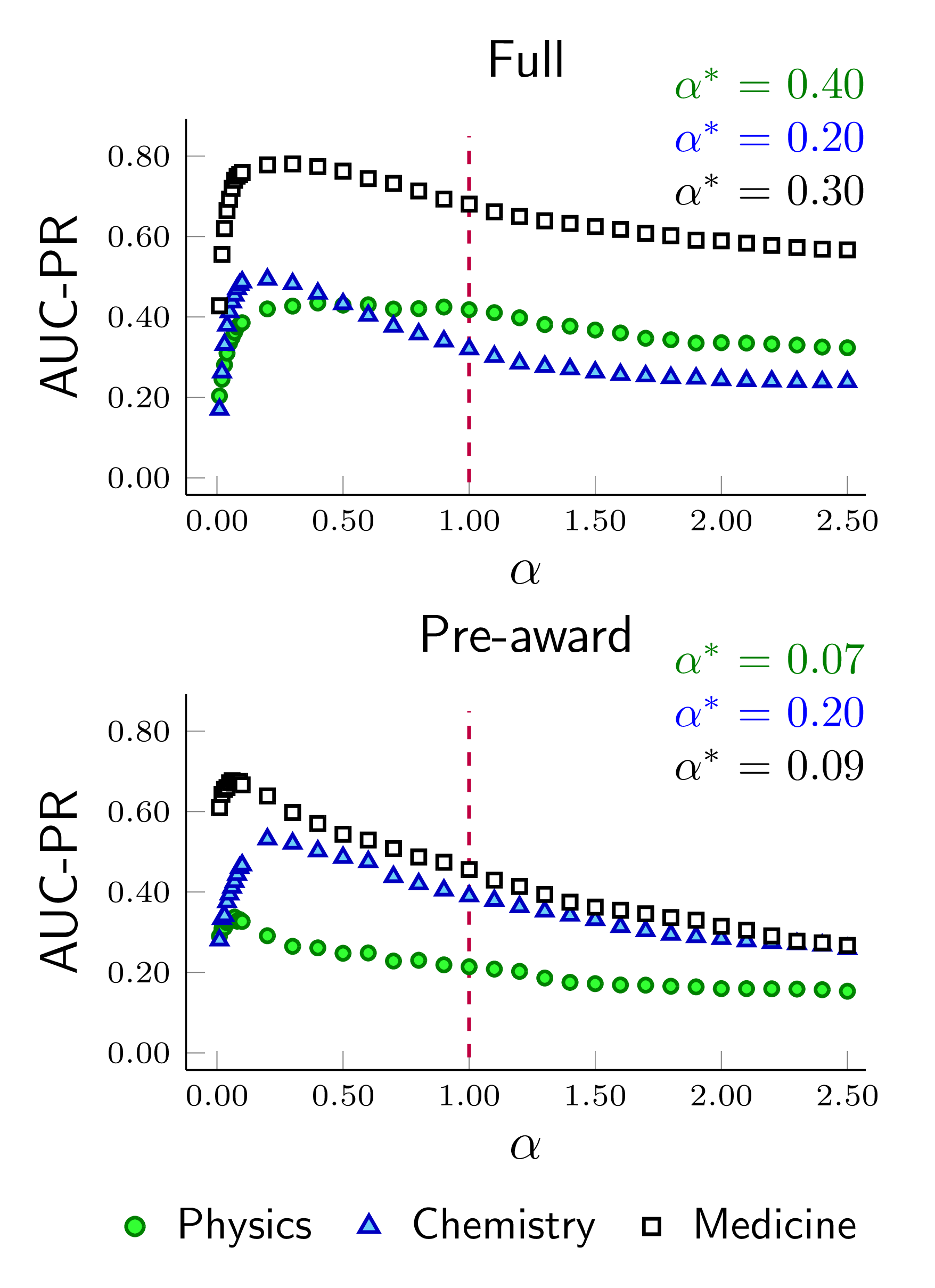}   
    \caption{Classification performance of $M_{\alpha}$ for varying $\alpha$. Different 
    symbols denote
    different fields. The dashed line $\alpha = 1$ separates the two regimes. We use the optimal values of $\alpha$ ($\alpha^\ast$) in our analyses.}
    \label{fig:alpha-pick}
\end{figure}

\noindent\para{Optimal $\alpha$ selection}
Recall that, unlike other measures, $M_{\alpha}$ has a tunable parameter $\alpha$. Therefore, for each task, we record the performance of $M_{\alpha}$ across a range of $\alpha$ values and plot the results in Fig.~\ref{fig:alpha-pick}.
We observe a slight dependence of the optimal $\alpha$-value ($\alpha^\ast$) on the task and the field. We use the corresponding $\alpha^\ast$ values while comparing the performance of $M_{\alpha}$ with other metrics. In each case, however, we find $\alpha^\ast < 1$,  which indicates that portfolios are most separable when the metric prioritizes consistent impact. 

We record the metrics' performance in Table~\ref{tab:combined-class}. In Appendix \ref{sec:appA}, we report the classification results on the American Physical Society (APS) bibliographic dataset.

Metrics agnostic to the distribution of citations appear to perform worse than their counterparts across either task. This includes the total number of papers $N$, as well as total citations $C_\text{tot}$, and maximum citations $C_\text{max}$. 
We highlight the performance of three metrics: $N$, $C_\text{avg}$, and $C_\text{max}$.
$N$ is consistently the worst performer because it does not account for the impact, only volume.
$C_\text{avg}$ is among the top performers considering the whole portfolio. We believe that is partly due to the nature of the distributions observed in Fig.~\ref{fig:NC}, where the Nobelists are likely to accumulate higher than average citations over their careers. 
However, performance for the pre-award portfolios is a bit worse, probably because we only consider the pre-award period of their careers. Winning the Prize has been shown to provide a tangible boost to the overall visibility of a scientist, resulting in more citations~\cite{inhaber1976quality}. 
The number of citations of the most cited paper $C_\text{max}$ is among the worst performers, which suggests that the one big-hit portfolio is not typical among Nobelists. This finding supports that scientists win the Nobel Prize after years of consistent, high-quality work. 

\begin{figure*}
    \centering
    \includegraphics[width=0.76\textwidth]{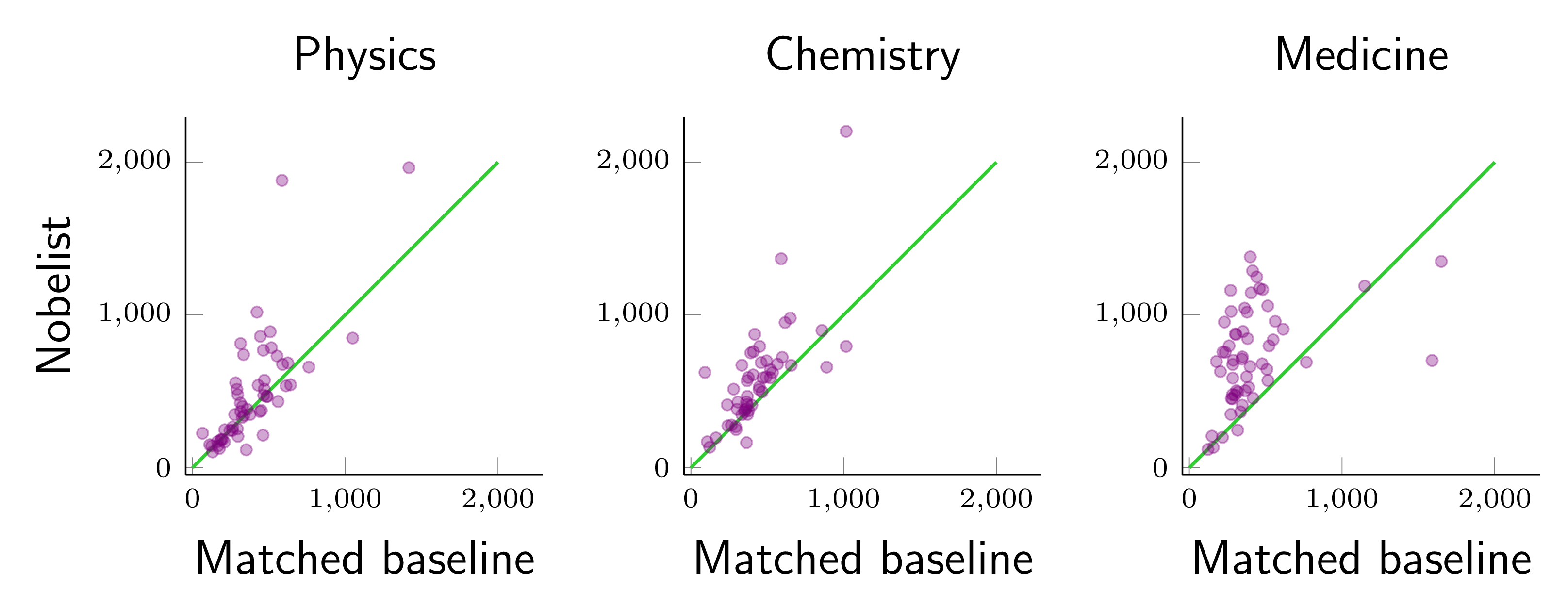}
    \caption{$E$-index of Nobelists versus baseline scientists with comparable number of papers and citations. We see a prevalent trend towards larger $E$ values for Nobelists. 58.2\% (Physics), 86.3\% (Chemistry), and 87.5\% (Medicine) of Nobelists have larger $E$ values than their counterparts.}
    \label{fig:matching}
\end{figure*}

We now shift our focus to the other category of indicators, \ie, ones sensitive to the citation distributions. We find that $H$ records mediocre performance despite rewarding consistency. Its dependence on productivity likely fails to account for the Nobelists with a few highly cited papers.
The $Q$-index performs poorly. However, its variant, $\tilde Q$, fares considerably better, which is consistent with the fact that it is similar to $M_{\alpha}$ for small $\alpha$. 

$M_{\alpha}$ and $E$ consistently rank in the top 2 positions. This further supports the hypothesis that Nobelists set themselves apart by producing a steady stream of high-impact work.

\begin{table}[t]
    \caption{AUC-PR values for the Full and Pre-award (PA) portfolio classification tasks.  The best-performing metrics for each field are marked in boldface. $M_\alpha$ and $E$ are the standout performers. Note that values across columns are not comparable as the baseline values are determined by the respective class imbalance ratios.}
    \centering
    \begin{NiceTabular}{@{} llcclcclc @{}}[colortbl-like]
    \toprule 
    \multirow{2}{*}{\textbf{Metric}}            & \multicolumn{2}{c}{\textbf{Physics}}   & \phantom{} & \multicolumn{2}{c}{\textbf{Chemistry}} & \phantom{} & \multicolumn{2}{c}{\textbf{Medicine}}  \\
    \cmidrule(lr){2-3} \cmidrule(lr){5-6} \cmidrule{8-9}
                   & Full & PA &  & Full & PA &  & Full & PA \\ \midrule 
    $N$            & 0.03 & 0.07      &  & 0.13 & 0.12      &  & 0.06 & 0.06      \\
    $C_\text{tot}$ & 0.21 & 0.15      &  & 0.43 & 0.34      &  & 0.52 & 0.24      \\
    $C_\text{avg}$ & 0.42 & 0.19      &  & 0.32 & 0.39      &  & 0.68 & 0.46      \\
    $C_\text{max}$ & 0.24 & 0.12      &  & 0.25 & 0.21      &  & 0.49 & 0.18      \\
    $H$            & 0.12 & 0.16      &  & 0.44 & 0.36      &  & 0.50 & 0.24      \\
    $G$            & 0.15 & 0.15      &  & 0.41 & 0.33      &  & 0.48 & 0.17      \\
    $\Tilde Q$     & 0.30 & 0.19      &  & 0.32 & 0.41      &  & 0.67 & 0.48      \\
    $Q$            & 0.08 & 0.15      &  & 0.13 & 0.20      &  & 0.26 & 0.45      \\
    $M_{\alpha}$ & 0.43          & \textbf{0.34} &             & 0.49          & \textbf{0.53} &             & \textbf{0.78} & \textbf{0.68} \\
    $E$          & \textbf{0.44} & 0.23          &             & \textbf{0.53} & 0.45          &             & 0.75          & 0.44          \\ 
    \bottomrule 
    \end{NiceTabular}
    \label{tab:combined-class}
\end{table}

\subsection{Identifying future Nobelists}

As a test of the predictive power of the metrics, we check whether we can identify scholars who received the Nobel Prize from 2018 to 2022, \ie, the period not covered by our WoS dataset.
First, we note that our set of baseline scientists may be missing some of these new Nobelists, in which case we add them manually, provided they have a GS profile. 

Then, for each metric, we construct a top 20 list of baseline scientists by ranking them in descending order and highlighting the Nobelists. We report the table for the $E$-index in the main text (Table~\ref{table:top20_entropy}), while the remaining lists can be found in Appendix \ref{sec:appC}. 

In Table~\ref{table:top20identified}, we show how many Nobelists appeared in the top 20 lists for each metric. $E$-index outperforms all other indicators, proving particularly effective for Medicine.

\begin{table}[b]
    \caption{Top 20 baseline scholars with the largest $E$-index in each discipline. The ones marked in boldface received the Nobel Prize between 2018 and 2022. Some authors are assigned multiple labels, so they may appear in multiple lists.}
    \label{table:top20_entropy}
    \resizebox{0.49\textwidth}{!}{
    \centering
    \begin{tabular}{@{}>{\columncolor{white}[0pt][\tabcolsep]}  ll ll >{\columncolor{white}[\tabcolsep][-10pt]}l @{}}
    \toprule
    \textbf{Rank} & \phantom{a} & \textbf{Physics} & \textbf{Chemistry} & \textbf{Medicine}\\ 
    \midrule
    1 && H. Dai & H. Dai & S. Kumar \\
    2 && A. L. Barabási & J. Godwin & R. A. Larson \\
    3 && D. Finkbeiner & R. Ruoff & A. L. Barabási \\
    4 && P. McEuen & K. L. Kelly & \textbf{G. L. Semenza} \\
    5 && I. Bloch & H. Wang & A. S. Levey \\
    6 && \textbf{A. Ashkin} & M. Egholm & \textbf{S. Paabo} \\
    7 && U. Seljak & L. Umayam & R. A. North \\
    8 && S. Inouye & L. Zhang & \textbf{A. Patapoutian} \\
    9 && \textbf{S. Manabe} & R. Freeman & J. Goldberger \\
    10 && M. Tegmark & P. Cieplak & M. Snyder \\
    11 && J. R. Heath & G. Church & J. Magee \\
    12 && L. Verde & \textbf{D. Macmillan} & \textbf{M. Houghton} \\
    13 && S. G. Louie & \textbf{G. Winter} & G. Loewenstein \\
    14 && D. I. Schuster & J. Kuriyan & S. Via \\
    15 && N. D. Lang & J. R. Heath & R. Jaeschke \\
    16 && B. Hammer & E. H. Schroeter & G. Hollopeter \\
    17 && D. Holmgren & W. Lin & S. J. Wagner \\
    18 && M. Lazzeri & W. L. Jorgensen & V. V. Fokin \\
    19 && L. P. Kouwenhoven & J. Clardy & \textbf{J. Allison} \\
    20 && M. Buttiker & D. Zhao & B. Moss \\
    \bottomrule
    \end{tabular}}
\end{table}

To further corroborate this conclusion, we matched each Nobelist with a baseline scientist with (nearly) identical $N$ and $C_\text{tot}$ values. In Fig.~\ref{fig:matching}, we plot the $E$-index of each Nobelist and matched baseline pair. 
We find that the $E$-index of Nobelists usually exceeds that of their matches. Some exceptions correspond to Nobelists with a low number of highly cited papers. Other outliers might be prominent scholars who have not yet received the award but might receive it in the future.

\begin{table}[htb]
    \caption{Count of Nobelists awarded in the period [2018, 2022] identified in the top 20 lists of various metrics. The numbers in parentheses indicate how many such Nobelists have a GS profile.}
    \label{table:top20identified}
    \centering 
    
     \begin{NiceTabular}{@{}  ll cc c @{}}[colortbl-like]
        \toprule
        \textbf{Metric} & \phantom{a} & \textbf{Physics} (9) & \textbf{Chemistry} (8) & \textbf{Medicine} (5) \\
        \midrule

        $N$ && 1 & 0 & 0 \\
        $C_\text{tot}$ && \textbf{2} & 0 & 3 \\
        $C_\text{avg}$ && 1 & 0 & 3 \\
        $C_\text{max}$ && 0 & 0 & 2 \\
        $H$ && \textbf{2} & 1 & 2 \\
        $G$ && \textbf{2} & 0 & 3 \\
        $\tilde Q$ && 1 & 1 & 3 \\
        $Q$ && 0 & 1 & 1 \\
        $M_{\alpha}$ && 1 & 0 & 2 \\
        $E$ && \textbf{2} & \textbf{2} & \textbf{5} \\
        \bottomrule
    \end{NiceTabular}
\end{table}

\section{Conclusion} \label{sec:conclusion}

In this work, we searched for productivity patterns in excellent scientific careers.
Specifically, we aimed to assess whether the output of high-profile scientists is more likely to be characterized by a low number of hit papers or by a consistent production of high-quality work. To address this question, we have examined the scientific portfolios of Nobel Prize winners in Physics, Chemistry, and Physiology or Medicine and checked which citation-based metrics are most suitable to recognize them among a much larger number of baseline scholars. In addition, we introduced two new metrics, the $E$-index and $M_\alpha$, that reward both consistency and high average impact (when $\alpha <1$).  

We found that the best-performing metrics are the ones that peak when citations are distributed among a considerable number of works rather than being concentrated on a few hit papers.
The $E$-index, in particular, proves especially effective in identifying future Nobelists. A portal for the calculation of $E$-index and other scores of individual performance can be found at \href{https://e-index.net/}{e-index.net}.

While there are Nobelists whose success relied on isolated hit papers, the most successful scientists usually stayed on top of their game for most of their careers.

\acknowledgments{This project was partially supported by grants from the Army Research Office (\#W911NF-21-1-0194) and the Air Force Office of Scientific Research (\#FA9550-19-1-0391, \#FA9550-19-1-0354). We acknowledge Aditya Tandon's help in this study's initial phase. This work uses Web of Science data by Clarivate Analytics provided by the Indiana University Network Science Institute and the Cyberinfrastructure for Network Science Center at Indiana University.}

\section*{Data Availability}
The data for Nobel laureates is available at~\cite{li2019dataset}. The disambiguated APS dataset is available at~\cite{sinatra2016quantifying}. The raw dataset for the APS can be requested at \href{https://journals.aps.org/datasets}{https://journals.aps.org/datasets}. The code is available at \href{https://github.com/siragerkol/Consistency-pays-off-in-science}{https://github.com/siragerkol/Consistency-pays-off-in-science}. Web of Science data is not publicly available.

\bibliographystyle{ieeetr}

\newpage
\appendix

\section{Analysis of APS data} \label{sec:appA}
In Table~\ref{table:aps}, we report the performance of all metrics for the classification task on the American Physical Society (APS) bibliographic dataset. The dataset used is the same as in~\cite{sinatra2016quantifying}. The dataset considers scientists whose career spans at least 20 years, have at least ten publications, and have authored at least one paper every five years. The portfolios of scientists consist of their publications until 2000 and citations received until 2010. The optimal $\alpha$ ($\alpha^\ast)$ for $M_{\alpha}$ is $0.6$.

\begin{table}[htb]
    \caption{AUC-PR values for the classification task in APS. The top performer is indicated in bold.}
    \label{table:aps}
    \centering 
    {
        \begin{NiceTabular}{@{} l c c @{}}[colortbl-like]
        \toprule
        \textbf{Metric} & \phantom{aaaaaaaaaaaaaaa} & \textbf{APS} \\
        \midrule
        $N$ && 0.02 \\
        $C_\text{tot}$ && 0.13 \\
        $C_\text{avg}$ && 0.16 \\
        $C_\text{max}$ && 0.10 \\
        $H$ && 0.14 \\
        $G$ &&  0.16  \\
        $\tilde Q$ && 0.27 \\
        $Q$ && 0.21 \\
        $M_{\alpha}$ && {\bf 0.28}   \\
        $E$ && 0.25  \\
        \bottomrule
    \end{NiceTabular}
    }
\end{table}

\section{Receiver Operating Characteristic Curve}
\label{sec:appB}

In Table~\ref{tab:rocauc}, we report the performances of metrics using ROC-AUC.

\begin{table}[!htb]
    \centering
    \caption{ROC-AUC values for the Full and Pre-award (PA) portfolio classification tasks. The best-performing metrics for each field are marked in boldface.}
    \begin{NiceTabular}{@{} llcclcclc @{}}[colortbl-like]
    \toprule 
    \multirow{2}{*}{\textbf{Metric}}            & \multicolumn{2}{c}{\textbf{Physics}}   & \phantom{} & \multicolumn{2}{c}{\textbf{Chemistry}} & \phantom{} & \multicolumn{2}{c}{\textbf{Medicine}}  \\
    \cmidrule(lr){2-3} \cmidrule(lr){5-6} \cmidrule{8-9}
                   & Full & PA &  & Full & PA &  & Full & PA \\ \midrule 
    $N$            & 0.68 & 0.57      &  & 0.88 & 0.70      &  & 0.78 & 0.51      \\
    $C_\text{tot}$ & 0.93 & 0.76      &  & \textbf{0.98} & 0.88      &  & 0.97 & 0.79      \\
    $C_\text{avg}$ & \textbf{0.96} & 0.83      &  & 0.97 & 0.91      &  & \textbf{0.99} & 0.88      \\
    $C_\text{max}$ & 0.95 & 0.75      &  & 0.97 & 0.84      &  & 0.98 & 0.76      \\
    $H$            & 0.86 & 0.75      &  & 0.96 & 0.86      &  & 0.94 & 0.78      \\
    $G$            & 0.86 & 0.76      &  & 0.96 & 0.84      &  & 0.92 & 0.73      \\
    $\Tilde Q$     & 0.94 & 0.81      &  & 0.95 & 0.89      &  & 0.98 & 0.89      \\
    $Q$            & 0.89 & 0.80      &  & 0.90 & 0.80      &  & 0.93 & 0.89      \\
    $M_{\alpha}$ & \textbf{0.96}          & \textbf{0.86} &             & \textbf{0.98}          & \textbf{0.94} &             & \textbf{0.99} & \textbf{0.93} \\
    $E$          & \textbf{0.96} & 0.83          &             & \textbf{0.98} & 0.93          &             & \textbf{0.99}          & 0.90          \\ 
    \bottomrule 
    \end{NiceTabular}
    
    \label{tab:rocauc}
\end{table}

\newpage

\section{Identification of recent Nobelists}
\label{sec:appC}
Similar to Table \ref{table:top20_entropy}, we report the top 20 baseline scientists for various indicators in Tables~\ref{table:top20_1} and~\ref{table:top20_2}. Scientists who have won the prize between 2018 and 2022 are marked in bold.

\begin{table*}[htb]
\caption{Top 20 baseline scholars with the largest $C_\text{tot}$, $C_\text{avg}$, $C_\text{max}$, and $H$ in each discipline.}
\label{table:top20_1}
\resizebox{0.95\textwidth}{!}{
\centering
\begin{tabular}{@{}>{\columncolor{white}[0pt][\tabcolsep]}  ll ll >{\columncolor{white}[\tabcolsep][-10pt]}l || ll ll >{\columncolor{white}[\tabcolsep][-10pt]}l @{}}
\toprule
&&&$C_\text{tot}$&&&&&$C_\text{avg}$&\\
\midrule
\textbf{Rank} & \phantom{a} & \textbf{Physics} & \textbf{Chemistry} & \textbf{Medicine} & \textbf{Rank} & \phantom{a} & \textbf{Physics} & \textbf{Chemistry} & \textbf{Medicine}\\ 
\midrule
1 && H. Dai & R. Ruoff & S. Kumar & 1 && S. Scherer & K. L. Kelly & S. Kumar \\
2 && A. L. Barabási & H. Dai & A. L. Barabási & 2 && D. Finkbeiner & P. McEwan & R. A. Larson \\
3 && A. Zunger & D. Zhao & \textbf{G. L. Semenza} & 3 && A. L. Barabási & L. Zhang & J. Jee \\
4 && S. G. Louie & W. L. Jorgensen & A. S. Levey & 4 && H. Dai & J. Godwin & A. L. Barabási \\
5 && M. Katsnelson & J. M. Tour & M. Snyder & 5 && J. H. Chen & L. Umayam & J. Goldberger \\
6 && D. Scott & K. Nicolaou & B. Moss & 6 && A. Rimmer & I. Bruno & A. Jánosi \\
7 && J. R. Heath & G. Church & J. Rouleau & 7 && M. Lazzeri & H. Dai & G. Hollopeter \\
8 && R. Car & E. W. Meijer & P. Henson & 8 && D. Holmgren & H. Wang & A. S. Levey \\
9 && \textbf{G. Parisi} & J. Hupp & J. Kanis & 9 && S. Inouye & E. H. Schroeter & P. A. Sirnes \\
10 && F. Guinea & J. R. Heath & J. McHutchison & 10 && D. I. Schuster & P. Cieplak & S. Via \\
11 && M. Scheffler & R. Car & P. McGeer & 11 && \textbf{A. Ashkin} & M. Egholm & \textbf{G. L. Semenza} \\
12 && F. S. Bates & C. B. Murray & J. D. Griffin & 12 && S. A. McCarthy & R. Ruoff & \textbf{A. Patapoutian} \\
13 && D. Finkbeiner & F. S. Bates & \textbf{J. Allison} & 13 && P. McEuen & J. H. Jensen & M. O'Donovan \\
14 && T. E. Mallouk & S. P. Nolan & A. Ascherio & 14 && L. Verde & R. Hawley & \textbf{S. Paabo} \\
15 && C. Stubbs & S. Marder & R. A. North & 15 && N. Peres & K. Dzepina & J. Kubota \\
16 && \textbf{A. Zeilinger} & R. Kaner & N. E. Davidson & 16 && M. Eppard & A. Hochbaum & J. Magee \\
17 && M. Tegmark & H. B. Schlegel & J. Bonventre & 17 && S. D. Bergin & S. C. Christiansen & R. Benediktsson \\
18 && D. R. Nelson & T. E. Mallouk & A. Phillips & 18 && C. Stewart & M. Wasa & M. Enge \\
19 && P. F. Michelson & J. Clardy & A. R. Tall & 19 && I. Bloch & R. Tsunashima & F. Cohen \\
20 && H. Sirringhaus & X. Chen & \textbf{M. Houghton} & 20 && S. Gronstal & R. Freeman & R. Ivanhoe \\
\midrule
&&&$C_\text{max}$&&&&&$H$&\\
\midrule
\textbf{Rank} & \phantom{a} & \textbf{Physics} & \textbf{Chemistry} & \textbf{Medicine} & \textbf{Rank} & \phantom{a} & \textbf{Physics} & \textbf{Chemistry} & \textbf{Medicine}\\ 
\midrule
1 && A. L. Barabási & G. Grynkiewicz & S. Kumar & 1 && H. Dai & H. Dai & \textbf{G. L. Semenza} \\
2 && A. Zunger & W. L. Jorgensen & A. L. Barabási & 2 && S. G. Louie & D. Zhao & B. Moss \\
3 && M. Katsnelson & J. H. Jensen & A. S. Levey & 3 && A. Zunger & R. Ruoff & M. Snyder \\
4 && N. Peres & P. McEwan & V. V. Fokin & 4 && M. Scheffler & K. Nicolaou & J. D. Griffin \\
5 && F. Guinea & J. R. Heath & W. Wilson & 5 && T. E. Mallouk & J. M. Tour & P. Henson \\
6 && J. R. Heath & K. M. Merz Jr. & T. Bleck & 6 && F. S. Bates & G. Church & R. A. North \\
7 && D. Finkbeiner & P. Cieplak & \textbf{M. Houghton} & 7 && A. L. Barabási & J. Hupp & A. S. Levey \\
8 && C. Stubbs & G. Church & B. Davis & 8 && P. F. Michelson & E. W. Meijer & P. McGeer \\
9 && P. Sutton & D. Zhao & J. Rouleau & 9 && D. Scott & T. E. Mallouk & A. Ascherio \\
10 && P. Nugent & R. Ruoff & M. Snyder & 10 && \textbf{G. Parisi} & S. P. Nolan & \textbf{J. Allison} \\
11 && S. Scherer & R. Car & J. McHutchison & 11 && D. R. Nelson & W. L. Jorgensen & J. Kanis \\
12 && R. Car & V. V. Fokin & D. A. Schoenfeld & 12 && \textbf{A. Zeilinger} & F. S. Bates & J. Bonventre \\
13 && L. Verde & C. B. Murray & G. Lamas & 13 && J. R. Heath & J. Clardy & A. R. Tall \\
14 && S. Scandolo & B. H. Hong & J. C. Nicolau & 14 && M. Razzano & X. Chen & A. L. Barabási \\
15 && M. Lazzeri & K. L. Kelly & A. Jánosi & 15 && G. Tosti & D. R. Nelson & D. Bers \\
16 && J. M. Soler & R. Kaner & B. M. Mayosi & 16 && P. Mulvaney & S. Marder & A. Phillips \\
17 && W. L. Barnes & M. Valko & \textbf{G. L. Semenza} & 17 && H. J. Freund & J. L. Atwood & A. L. Warshaw \\
18 && J. Fabian & S. Miertus & P. Barter & 18 && R. V. Grondelle & W. Lin & J. Rouleau \\
19 && D. Holmgren & T. Gunnlaugsson & M. O'Donovan & 19 && P. Knight & \textbf{C. Bertozzi} & D. A. Schoenfeld \\
20 && A. A. Balandin & H. Q. N. Gunaratne & D. Kleiner & 20 && M. Katsnelson & J. R. Heath & H. B. El-Serag \\
\bottomrule
\end{tabular}}
\end{table*}

\begin{table*}[htb]
\caption{Top 20 baseline scholars with the largest $G$, $\tilde Q$, $Q$, and $M_{\alpha}$ in each discipline.}
\label{table:top20_2}
\resizebox{0.95\textwidth}{!}{
\centering
\begin{tabular}{@{}>{\columncolor{white}[0pt][\tabcolsep]}  ll ll >{\columncolor{white}[\tabcolsep][-10pt]}l || ll ll >{\columncolor{white}[\tabcolsep][-10pt]}l @{}}
\toprule
&&&$G$&&&&&$\tilde Q$&\\
\midrule
\textbf{Rank} & \phantom{a} & \textbf{Physics} & \textbf{Chemistry} & \textbf{Medicine} & \textbf{Rank} & \phantom{a} & \textbf{Physics} & \textbf{Chemistry} & \textbf{Medicine}\\ 
\midrule
1 && H. Dai & H. Dai & \textbf{G. L. Semenza} & 1 && D. I. Schuster & K. L. Kelly & R. A. Larson \\
2 && A. L. Barabási & R. Ruoff & A. S. Levey & 2 && H. Dai & L. Umayam & G. Hollopeter \\
3 && S. G. Louie & W. L. Jorgensen & A. L. Barabási & 3 && C. Bharucha & \textbf{D. Macmillan} & G. Gerdeman \\
4 && M. Katsnelson & D. Zhao & M. Snyder & 4 && D. Finkbeiner & J. B. Binder & J. Kubota \\
5 && A. Zunger & G. Church & J. Rouleau & 5 && A. Paic & E. H. Schroeter & S. J. Wagner \\
6 && J. R. Heath & J. M. Tour & B. Moss & 6 && S. B. Ogale & S. Furyk & C. Schmaltz \\
7 && R. Car & J. R. Heath & P. Henson & 7 && S. Inouye & L. Zhang & \textbf{A. Patapoutian} \\
8 && D. Scott & R. Car & J. McHutchison & 8 && \textbf{S. Manabe} & S. C. Christiansen & J. Magee \\
9 && F. Guinea & E. W. Meijer & \textbf{J. Allison} & 9 && P. McEuen & H. Dai & S. Koh \\
10 && \textbf{G. Parisi} & C. B. Murray & J. D. Griffin & 10 && S. Gronstal & A. Hochbaum & S. Via \\
11 && C. Stubbs & J. Hupp & A. Ascherio & 11 && N. D. Lang & V. Bowry & M. Enge \\
12 && F. S. Bates & K. Nicolaou & P. McGeer & 12 && P. Y. Huet & M. Wasa & R. A. North \\
13 && M. Tegmark & R. Kaner & J. Kanis & 13 && A. Zupan & T. Arai & \textbf{S. Paabo} \\
14 && T. E. Mallouk & F. S. Bates & N. E. Davidson & 14 && B. S. Chandrasekhar & A. Zulys & M. B. Olofsson \\
15 && \textbf{A. Zeilinger} & H. B. Schlegel & R. A. North & 15 && W. Borghols & I. Bruno & \textbf{G. L. Semenza} \\
16 && L. Verde & T. E. Mallouk & \textbf{M. Houghton} & 16 && U. Seljak & J. Farr & A. L. Barabási \\
17 && D. R. Nelson & W. Lin & A. R. Tall & 17 && I. Bloch & R. Slone & J. Z. Long \\
18 && H. Sirringhaus & S. Marder & J. Bonventre & 18 && A. L. Barabási & K. Dzepina & C. Zhang \\
19 && M. Scheffler & D. R. Nelson & H. Y. Chang & 19 && C. Stewart & C. S. Day & V. V. Fokin \\
20 && P. Corkum & S. P. Nolan & H. B. El-Serag & 20 && C. Gross & F. Pesciaioli & S. Hamilton \\
\midrule
&&&$Q$&&&&&$M_{\alpha}$&\\
\midrule
\textbf{Rank} & \phantom{a} & \textbf{Physics} & \textbf{Chemistry} & \textbf{Medicine} & \textbf{Rank} & \phantom{a} & \textbf{Physics} & \textbf{Chemistry} & \textbf{Medicine}\\ 
\midrule
1 && S. Wiedmann & D. Ravelli & S. Ripke & 1 && D. Finkbeiner & L. Umayam & R. A. Larson \\
2 && S. Guragain & T. K. Sen & M. Enge & 2 && H. Dai & K. L. Kelly & G. Hollopeter \\
3 && M. Lahaye & F. Pesciaioli & A. Chaix & 3 && C. Bharucha & L. Zhang & S. Via \\
4 && M. Schwartz & G. F. Schneider & A. Klepper & 4 && S. Inouye & T. Arai & M. Enge \\
5 && S. Faez & M. Bodnarchuk & K. Ihle & 5 && S. Gronstal & R. Freeman & S. J. Wagner \\
6 && H. Min & S. Sreejith & R. A. Larson & 6 && D. I. Schuster & H. Dai & J. Kubota \\
7 && D. I. Schuster & A. M. Spokoyny & J. Rooney & 7 && J. H. Chen & C. S. Day & A. L. Barabási \\
8 && D. Lomidze & W. Borghols & A. K. Singh & 8 && \textbf{S. Manabe} & Q. Jia & S. Kumar \\
9 && K. Theofilatos & A. B. D. Nandiyanto & J. Goldberger & 9 && C. Gross & W. Borghols & S. Koh \\
10 && H. Rouault & T. Pellegrino & G. Gerdeman & 10 && S. B. Ogale & J. Clardy & J. Magee \\
11 && H. B. Akkerman & E. Bakota & G. Hollopeter & 11 && P. Y. Huet & H. Wang & \textbf{S. Paabo} \\
12 && J. H. Chen & M. Raj & M. V. B. Malachias & 12 && L. Winoto & P. B. Armentrout & \textbf{A. Patapoutian} \\
13 && M. R. Becker & Y. Vasquez & D. Ono & 13 && W. Borghols & D. Zhao & R. A. North \\
14 && C. Ward & Z. Chen & A. Rungatscher & 14 && P. McEuen & R. Tsunashima & M. B. Olofsson \\
15 && W. Borghols & A. Datar & D. Mertz & 15 && U. Seljak & F. Ibarra & S. Krauss \\
16 && V. Simonyan & H. Wang & Y. Xue & 16 && N. D. Lang & E. H. Schroeter & T. Maack \\
17 && D. Finkbeiner & N. Lee & Y. Gao & 17 && B. S. Chandrasekhar & M. Wasa & S. Hamilton \\
18 && M. Eichenfield & \textbf{D. Macmillan} & \textbf{A. Patapoutian} & 18 && C. Sanner & T. Rockway & J. Goldberger \\
19 && J. C. Brant & A. Hochbaum & T. Mustakov & 19 && A. Paic & J. Farr & A. S. Levey \\
20 && J. L. Niedziela & L. Umayam & F. S. Regateiro & 20 && I. Bloch & X. Periole & H. L. Wang \\
\bottomrule
\end{tabular}}
\end{table*}

\clearpage

\end{document}